\documentclass[useAMS,usenatbib]{mn2e}

\usepackage{natbib}
\usepackage{graphicx}
\usepackage{times}

\usepackage{xcolor}

\title{A Recent Major Merger Tale for the Closest Giant Elliptical Galaxy Centaurus A} 

\author[Jianling Wang et al.]{ 
  Jianling Wang$^{1}$\thanks{E-mail:wjianl@bao.ac.cn}, 
  Francois Hammer$^{2}$\thanks{E-mail:francois.hammer@obspm.fr}, 
  Marina Rejkuba$^{3}$,
  Denija Crnojevi\'{c}$^{4}$,
  Yanbin Yang$^{2}$\\
{$^1$ CAS Key Laboratory of Optical Astronomy, National Astronomical Observatories, Beijing 100101, China,} \\
{$^2$ GEPI, Observatoire de Paris, CNRS, Place Jules Janssen 92195, Meudon, France.}\\
{$^{3}$European Southern Observatory, Karl-Schwarzschild-Stra\ss e 2, D-85748 Garching bei M\"unchen, Germany}\\
{$^{4}$University of Tampa, 401 West Kennedy Boulevard, Tampa, FL 33606, USA}\\
 }

\begin{document} 

\date{Received ; accepted}

\maketitle

\begin{abstract}

We have used hydrodynamical simulations to model the formation of the closest giant elliptical
galaxy, Centaurus A. We find that a single major merger event with a mass ratio up to 1.5, and which has happened $\sim 2$ Gyr ago, is able to reproduce many of its properties, including galaxy kinematics, the inner gas disk, stellar halo ages and metallicities, and numerous faint features observed in the halo.
The elongated halo shape is mostly made of
progenitor residuals deposited by the merger, which also contribute to stellar shells observed in the Centaurus A halo. The current model
also reproduces the measured Planetary Nebulae line of sight velocity and their velocity dispersion. Models with small mass
ratio and relatively low gas fraction result in a de Vaucouleurs profile
distribution, which is consistent with observations and model expectations. A recent merger 
left imprints in the age distribution that are consistent with the 
young stellar and Globular Cluster populations (2-4 Gyrs) found within the halo. We conclude that even if not all properties of Centaurus A have been accurately reproduced, a recent major merger has likely occurred to form the Centaurus A galaxy as we observe it at present-day.

\end{abstract}

\begin{keywords}
 Galaxies: evolution - Galaxies: interactions - individual: NGC 5128
\end{keywords}

\section{Introduction}

Understanding the star formation history and mass assembly history of galaxies 
is one of the most challenging tasks in modern astrophysics. Elliptical galaxies are the most massive ones in the classification of Hubble \citep{Hubble1936}. They contain $\sim 22$\% of the total mass in stars in the local universe and this fraction goes to 75\% for spheroids, including S0 and spiral bulges. The massive Early Type Galaxies (ETG) are believed to form according to a two phase scenario  \citep[e.g.][]{Daddi2005,Oser2010,Arnold2011,Naab2017,Iodice2017,Pulsoni2020}. At high redshift, gas
collapses in the centre of dark matter and intense star formation occurs,
to be subequently quickly quenched. Then in a second stage, the accretion is dominant and ETGs grow efficiently in size through a series of mergers. The complex formation process through mergers and accretion of nearby galaxies make it difficult to disentangle mixed material and stars originating from different progenitors. 

Centaurus A (Cen A, NGC 5128), the central galaxy of the Centaurus group, is the closest easily observable giant elliptical 
galaxy with a distance of 3.8 Mpc \citep{Harris2010}. The iconic optical image of Cen A with a prominent twisted dust lane traversing a large spheroid has led to galaxy classification as peculiar elliptical or sometimes also S0 or S0p \citep[see][for a more details about Cen A's classification]{Harris2010b}. This morphology has been ascribed already in the 50-ies to a possible merger origin. \citet{Baade1954} interpreted the object as consisting of two nebulae, an elliptical nebula and a second system, possibly a spiral, that are {\it "in a state of strong gravitational interaction, perhaps actually in collision."} A number of studies have pointed out that the main body of the galaxy has a light distribution of a "normal" elliptical following closely the de Vaucouleur's $r^{1/4}$ profile \citep{Sersic1958,vandenbergh1976,Dufour1979}, and that the denomination as peculiar was perhaps undeserving and only a result of the proximity, that offers more detailed observations than possible in more distant systems \citep{Ebneter1983, Harris2010b}. The proximity of Cen A provides excellent opportunities to study the galaxy in exquisite detail, and a possibility to disentangle its formation history. 

\citet{Wang2012, Wang2015} and \citet{Hammer2010,Hammer2018} modeled the formation of nearby giant spirals using numerical simulations of major mergers of two gas rich spiral galaxies. Such a code can be adapted to explore a possibility of reproducing the observed properties of Cen A as a result of a major merger. Through comparison of observational properties of Cen A with those predicted by merger simulations we can learn about galaxy formation processes. 

In spite of already mentioned ideas about the merger origin of Cen A, some of its properties are suggesting an early rapid formation with subsequent evolution that included accretion(s) of satellite galaxies. This is in particular supported by the observed age distribution of stars \citep{Rejkuba2005, Rejkuba2011} and globular clusters (GCs) \citep{Kaviraj2005, Beasley2008, Woodley2007, Woodley2010} that contain a bulk of the population that formed $> 10$~Gyr ago, and up to 20-30\% that formed later having ages as young as $\sim 2-4$~Gyr. \citet{Beasley2003} have compared the stellar halo and GCs metallicity distribution functions with predictions of a $\Lambda$ cold dark matter ($\Lambda$CDM) semi-analytic galaxy formation model, finding that the vast majority of star formation in the model occurs quiescently, while the red metal-rich GCs require hierarchical mergers, leading to predictions for their age-metallicity distribution. The warped disk that crosses the centre of Cen A and the presence of HI and CO in the disk and in shells surrounding the galaxy have been interpreted as coming from a relatively recent minor merger with a gas rich spiral akin in size to M33 \citep{Quillen1992, Mirabel1999, Charmandaris2000}. 

The formation scenario in which Cen A has experienced a relatively recent major merger has been invoked in the literature due to its perturbed morphology with the central dust lane, filaments and shells, and it could perhaps also explain the presence of the active galactic nucleus (AGN) in its centre \citep[see the review by ][for a historical perspective and further references]{Israel1998}.   \citet{Peng2004GCS} suggested that the GC System and Planetary Nebulae data in Cen A support a scenario according to which the main body of the galaxy was formed several Gyr ago through a dissipational merger of two unequal-mass disk galaxies and continued to grow through accretion of further satellites. \citet{Mathieu1996} built a triaxial dynamical model interpreting the Cen A Planetary Nebulae kinematics measured by \citet{Hui1995} as the results of a major merger with 3:1 mass ratio. Numerical simulations in which an elliptical galaxy is formed through a major merger of spiral galaxies have been made by \citet{Bekki2003} as well as by \citet{Bekki2006}. They were used to explain the halo metallicity distribution and PN kinematics, respectively. Additional simulations that include gas dissipation and star formation are necessary to make further progress \citep{Peng2004PN}. It has been shown that realistic mergers require a correct treatment of the gas through hydrodynamical simulations, and by using consistent hydrodynamical solvers (see \citealt{Hopkins2015}, and references therein).

\citet{Hammer2013} proposed that the vast thin disk of satellites around our closest giant spiral M31 \citep{Ibata2013} could be linked to a major merger. The model of M31 \citep{Hammer2018} as a recent 2-3 Gyr old major merger reproduced successfully the giant stream and halo substructures in M31 \citep{Ibata2007}. This prompts the question of whether a similar explanation could be applied also to Cen A.

The aim of this paper is to verify if a numerical simulation of a major merger could explain several observational properties in Cen A. In Section 2 we describe data used for comparison with the models. The simulation method and initial conditions are described in Section 3. Results including comparisons to observations are presented in Section 4. In the last section, we discuss and summarize our results.

\section{THE PROPERTIES OF CENTAURUS A}\label{sec:property}

Centaurus A is one of the nearest and largest radio galaxies, and its optical counterpart is the giant elliptical galaxy NGC 5128 with the total integrated magnitude $M_B = -21.2$~mag \citep{Dufour1979}. Throughout the paper, while we examine mostly the optical properties of the elliptical galaxy, we use its radio source name abbreviated to Cen A. Based on a review of distance measurements, \citet{Harris2010} derived the best-estimate distance for NGC~5128 of $3.8 \pm 0.1$~Mpc\footnote{At the 3.8~Mpc distance 1\arcmin\, is equivalent to $1.1$~kpc}. The integrated light measurements in the inner parts of the galaxy revealed a luminosity distribution similar to that of an E2 galaxy type following a de Vaucouleur's law over the range $20 < \mu_B< 25$~mag arcsec$^{-2}$ and having an effective radius of $R_{\mathrm{eff}}=305\arcsec$ measured in $B$-band \citep{Dufour1979}. At the distance of 3.8~Mpc this implies $R_{\mathrm{eff}}=5.6$~kpc.   

\citet{Fall2018} have derived the stellar mass of Cen A based on the observed $B - V$ colors and the predicted relation between $M_{\star}/L_K$ and $B-V$ from stellar population models, assuming $M_{\star}/L_K = 0.8$. The $K$ band luminosity from Two Micron All Sky Survey (2MASS) survey \citep{Skrutskie2006} has been corrected for the bias that underestimates luminosity for large galaxies by their "aperture corrections" \citep{Romanowsky2012}. After re-scaling the distance to 3.8 Mpc and the Initial Mass Function (IMF) to a 
Salpeter 'diet', the total stellar mass is estimated to be $\sim 2\times 10^{11}$ M$_{\odot}$.

The total dynamical mass has been measured using GCs, PN, HI gas, X-ray emission, and satellites. While these tracers cover different distances, the total mass is between $2.2 \times 10^{11}$~M$_{\odot}$ at $\sim 15$~kpc \citep[][corrected to 3.8~Mpc distance]{Schiminovich1994} and $1.1 \times 10^{12}$~M$_{\odot}$ at $\sim 80$~kpc \citep{Woodley2010kinematic}. Considering also dynamics of satellites, the mass of the whole Centaurus group is $9 \times 10^{12}$~M$_{\odot}$ \citep{Woodley2006, Karachentsev2007}. The $M/L_B$ ratio is lower than expected for an elliptical with values between 7--15 within inner 20~kpc and increasing outwards reaching $M/L_B=52 \pm 22$ at 45~kpc in the halo and $M/L_B=153 \pm 50$ for the Centaurus group  \citep{Kraft2003, Woodley2006, Samurovic2006}.

The giant elliptical is traversed by a warped dust lane \citep{Quillen2006} surrounded by a stellar ring populated by young red supergiant stars \citep{Kainulainen2009}, and it hosts a system of shells \citep{Malin1978, Peng2002} and an extended halo with numerous low surface brightness streams \citep{Crnojevic2016}. 

Most of the gas in Cen A is located in a disk that follows the dust lane in the
centre of the galaxy \citep{vanGorkom1990, Eckart1990, Quillen1992,
Espada2019}. Furthermore, HI and molecular gas have also been detected
extending out to about 15 kpc in the halo of Cen A \citep{Schiminovich1994,
Charmandaris2000, Oosterloo2005}. HI in the central area is in close to edge-on
structure along the central dust lane that is roughly perpendicular to the jet,
which is along PA$=50^\circ$ axis \citep{Tingay1998}. The central HI,
CO and dust are all part of a coherent warp that extends between 2 to 6500 pc \citep[see][for a review]{Quillen2010}.
There are multiple folds in this warped disk.  
The northern radio lobe is likely pointing towards us, while the southern lobe
points away. The gas and dust disk is mostly visible in the optical in the
north-eastern part where it is in front of the galaxy, while in the
south-western side it is partly obscured by the galaxy body
\citep{Morganti2010, Struve2010}.

The gas in the halo is distributed in a ring-like structure that rotates in the same direction as the main stellar body of the galaxy, and molecular gas is found to be located in close proximity of stellar shells \citep{Charmandaris2000}. In the north-eastern region of the HI ring, along the radio jet direction, there is evidence of recent star formation \citep{Graham1998, Mould2000, Rejkuba2001, Rejkuba2002} possibly triggered by interaction of the jet with the interstellar medium  \citep{Oosterloo2005, Salome2016, Santoro2016}. The total HI mass of $4.9 \times 10^8$~M$_{\odot}$ in the disk and another $\sim 5 \times 10^7$~M$_{\odot}$ in the shells were reported by \citet{Schiminovich1994, Struve2010}.  A similar amount of molecular gas was found by \citet{Charmandaris2000, Wild2000, Salome2016APEX}. The overall HI fraction in Cen A is relatively low for an ETG with $M_{HI} /L_B = 0.01$ \citep{Struve2010}. \citet{Parkin2012} reported a total dust mass of $(1.59 \pm 0.05) \times 10^7$~M$_{\odot}$ and total gas mass of $(2.7 \pm 0.2) \times 10^9$~M$_{\odot}$ based on Herschel and JCMT observations. Most recent high sensitivity and high spatial resolution maps of the central disk in Cen A made in CO (1-0) with ALMA \citep{Espada2019} found even larger reservoir of molecular gas amounting to $1.6 \times 10^9$~M$_{\odot}$.


Thanks to its proximity, the stellar halo of Cen A can be resolved into individual red giant branch (RGB) stars and from their colour distribution it is possible to measure a metallicity distribution function (MDF), in addition to surface brightness distribution and mean metallicity gradient. The stellar MDF has been derived from optical photometry obtained with the HST in fields ranging in distance between 8 -- 140~kpc \citep{Harris1999, Harris2000, Harris2002, Rejkuba2005, Rejkuba2014} from the centre of Cen A. Stellar halo age distribution was derived from the deepest HST field located $\sim 40$~kpc south of the centre of the Cen A. Two burst models with 70-80\% of the stars forming in a short burst $12 \pm 1$~Gyr ago, and the 20-30\% of the stars forming 2--4 Gyr ago, provided the best fit to these observations \citep{Rejkuba2011}. This age distribution is consistent with the globular cluster age distribution based on $U-B$ photometry \citep{Kaviraj2005} and on low-resolution spectroscopy \citep{Peng2004GCS, Beasley2008, Woodley2010}. Wider area imagers, VIMOS on the VLT and Megacam on the Magellan telescope, were used to trace the surface density and metallicity gradients in the outer halo \citep{Bird2015, Crnojevic2013, Crnojevic2016}. The resolved stellar halo studies showed a relatively shallow metallicity gradient with a slope $\Delta\mathrm{[M/H]}/\Delta R = - 0.0054 \pm 0.0006$~dex kpc$^{-1}$, or $\Delta\mathrm{[M/H]}/\Delta R_{\mathrm{eff}} = - 0.030 \pm 0.003$~dex per $R_{\mathrm{eff}}$ \citep{Rejkuba2014}. This may indicate that bulk of the halo was not assembled through accretion of many low mass satellites, but rather from few, massive ones.

Over a thousand Planetary Nebulae \citep[PN,][]{Hui1995, Peng2004PN, Walsh2015} and almost six hundred GCs \citep{Peng2004GCS, Beasley2008, Woodley2010kinematic} have measured velocities providing kinematic information from centre out to $\sim 10$~R$_{\mathrm{eff}}$, with a few confirmed clusters and PN as far out as 15.5~R$_{\mathrm{eff}}$ or 85~kpc \citep{Walsh2015}, and many more cluster candidates out to 150~kpc \citep{Taylor2017, voggel2020}. The halo of Cen A shows a disk-like feature, which has a large rotation along the major axis flattening at 100 km/s as traced by PN
\citep{Peng2004PN}. The zero-velocity contour of the velocity field is
perpendicular to the stellar major axis with a pronounced twist. 
GCs show different kinematics between metal-poor and metal-rich component \citep{Peng2004GCS, Woodley2010kinematic}. The metal-poor GCs are supported by dispersion ($149 \pm 4$~km/s) with a flat dispersion profile extending to 20\arcmin and then possibly increasing outwards. The metal-rich GCs have a similar velocity dispersion, but they also exhibit a rotation of $43 \pm 15$~km/s around the isophotal major axis. The globular cluster kinematics was discussed both as providing evidence supporting  \citep{Peng2004GCS} and against \citep{Woodley2010kinematic} a recent major merger formation scenario for the galaxy.

The studies mentioned above are mainly focusing on the large scale properties of the Cen A halo, its stellar population content and dynamics.  The properties and studies of the radio jet and black hole in the center region are beyond the scope of our analysis, and are not discussed in this paper.

\begin{table*}
\begin{center}
\caption{Parameters of the four models used in this study. The following parameters are listed in table rows: (1) the mass ratio of progenitor galaxies; (2--5) the initial angles (in degrees) of the progenitors with respect to the orbital plane; (6 and 7) initial gas fraction for the primary and the secondary progenitor; (8 and 9) initial scale length of the stellar components of each progenitor; (10 and 11) initial scale length of the gas component of each progenitor; (12) pericenter in kpc; (13) eccentricity orbital parameter; (14) number of particles; (15) particle mass ratio; (16) softening length; (17) the time that best matches the observations after beginning the simulations; (18 and 19) S{\'e}rsic index and effective radius used to fit the sufrace mass density; (20) The final gas fraction. (21) The final gas mass. }  
\begin{tabular}{lcccc}
\hline \hline
parameters                     & Model-6& Model-7&Model-10 & Model-11              \\
\hline
mass ratio                     & 1.0    & 1.0    &  1.0    &  1.5                 \\
Gal1 incy                      & 90     & 90     &  90     &   90                  \\
Gal1 incz                      &-90     &-90     & -90     &  -90                  \\
Gal2 incy                      & 70     & 80     &  80     &   80                  \\
Gal2 incz                      &-40     &-50     & -50     &  -50                  \\
Gal1 gas fraction              & 0.2    & 0.2    & 0.2     &  0.2                  \\
Gal2 gas fraction              & 0.2    & 0.2    & 0.4     &  0.4                  \\
Gal1 $h_{\rm star}$(kpc)       & 5.1    & 5.1    & 8.1     &  8.5                  \\
Gal2 $h_{\rm star}$(kpc)       & 5.1    & 5.1    & 5.1     &  4.86                 \\
Gal1 $h_{\rm gas}$(kpc)        &10.2    & 10.2   & 16.2    &  17.0                 \\
Gal2 $h_{\rm gas}$(kpc)        &10.2    & 10.2   & 15.3    &  14.58                \\
$r_{peri}$ (kpc)               & 20     & 20     &  18     &  25                   \\
eccentricity                   & 1.0    & 1.0    & 1.0     &  1.0                  \\
N$_{\rm particle}$             & 1.85M  & 1.85M  & 1.85M   &  1.85M                \\
m$_{dm}$:m$_{\rm star}$:m$_{\rm gas}$  &4:1:1   & 4:1:1  & 4:1:2   &  4:1:1                \\
softening($\epsilon_{\rm dm}$:$\epsilon_{\rm star}$:$\epsilon_{\rm gas}$)(kpc)& 0.3:0.1:0.1& 0.3:0.1:0.1&0.3:0.1:0.1&0.3:0.1:0.1  \\
\hline
Observed time (Gyr)            & 5.8    & 5.8    & 6.4     &  5.4                  \\
Sersic index($n$)              & 4.7    & 4.4    & 6.2     & 4.4                      \\
Effective radius (kpc)         & 3.7    & 4.0    & 4.4     & 6.3                   \\
Final gas fraction             & 3.2\%  & 3.5\%  & 4.1\%   &  8.8\%                 \\
Gas Mass (r$<50$ kpc) (10$^{10}$M$_{\odot}$)& 0.61  & 0.68    & 0.73  &  1.58      \\
\hline
\end{tabular}
\end{center}
\label{tab:par}
\end{table*}

\section{NUMERICAL SIMULATIONS AND INITIAL CONDITIONS}

The numerical simulations were carried out with GIZMO code \citep{Hopkins2015},
which is based on a new Lagrangian method for hydrodynamics, and has
simultaneously properties of both Smoothed Particle Hydrodynamics (SPH) and
grid-based/adaptive mesh refinement methods. We have implemented into GIZMO
star formation and feedback processes as described in \citet{Wang2012}  following the method of \citet{Cox2006}. The cooling process has been implemented in GIZMO with an updated version of \citet{Katz1996}. This code has been used in simulating the formation of Magellanic Stream \citep{Wang2019}.

The initial conditions were set up following the procedure of \citet{Wang2012,
Wang2015} and \citet{Hammer2010,Hammer2018}, who modeled the formation of a nearby giant spirals after a major merger of two gas rich spirals. Since Cen A is an early type galaxy with low gas fraction,
progenitors with moderate gas fractions ($20\%-40\%$) are used, which leads to a larger bulge fraction for the merger remnant \citep{Hopkins2009,Hopkins2010}. Moreover, \citet{Sauvaget2018} experimented with a series of major mergers showing that only those with an initially low gas fraction and low mass ratio ($<$2) are able to produce giant elliptical remnants. All the simulations are performed assuming a baryon fraction of 9\%  \citep{Wang2012,Wang2015}. The initial
gas disk scale-length is 2-3 times larger than that of the stellar disk, since
observations show that gas disks are more extended than stellar disks
\citep{Kruit2007}. 


There are several thin streams in the halo of Cen A that, if associated with a major merger, would favor a prograde orbit for one progenitor. Indeed, \citep{Toomre1972} showed that for prograde encounters (orbital angular momentum aligned with those of the initial disk galaxies), particles are more in resonance with the tidal field and result in far more prominent tidal tails than retrograde encounters (see also realizations of tidal tails by \citealt{Wang2012,Wang2015}). This also applies for interpreting the straight stream \citep{Crnojevic2016}, which leads us to use a prograde orbit for one of the progenitors.
We also use a polar orbit for the other progenitor, since
\citet{Bekki2006} found that a collision with a highly inclined orbital
configuration can reproduce Cen A's kinematics \citep{Peng2004PN}.
Observations show ongoing star formation with $\sim 15$~Myr old stars  along the jet direction in the north-east halo extending up to $\sim 20$~kpc \citep{Graham1998, Mould2000, Rejkuba2001}. Additional ionised gas filaments are observed up to 35~kpc distance\citep{Neff2015}. However, beyond those relatively confined areas with ongoing low efficiency star formation  \citep{Salome2016}, the stellar age distribution within $40$~kpc in the halo has a moderate ($20-30$\%) fraction of stars that formed as recently as 2~Gyr ago \citep{Rejkuba2011}.
This
epoch could indicate the fusion time of the merger \citep{Hammer2018}, during which some material can be easily ejected from the central regions. 
We compare the simulated merger remnants at the time that corresponds to 2$\pm$1 Gyr after fusion with observations.
The timescale for the current major merger is longer than that for minor merger models. \citet{Quillen1993} proposed a timescale of the order of 0.2 Gyr after accretion of a minor spiral to model the warp formation. This  is also roughly consistent with the shell-like feature formation timescale \citep{Schiminovich1994,Peng2002}. 
We developed an efficient software that enables three dimensional visualisation of results to model nearby galaxy formation.
For further details about this software we refer to \citet{Hammer2010,Hammer2018} and \citet{Wang2012,Wang2015}.

Following the same procedure as done in our previous studies \citep{Wang2012,Wang2015,Hammer2010,Hammer2018}, we have optimized the parameter space, including the pericenter, two inclination angles for each progenitor, the initial gas fractions and mass ratio. 
Given the large parameter space, It is impossible to examine all parameters in detail. However, the above constraints help reduce the parameter space significantly. During its exploration, we first build large coarse grid of parameters with values in the following ranges: (1) mass ratio of progenitors between 1 to 3, (2) the orbital parameters with pericenter distance between 5 kpc to 40 kpc and eccentricity ranging from 0.9 to 1, (3) different inclination angles of initial progenitors. Besides these, we also checked our major merger library that was built specifically for NGC 5907 and NGC 4013 \citep{Wang2012, Wang2015} in order to further constrain the parameter space. 
Snapshots from each simulation have been examined using 0.1 Gyr time steps. After finding possible candidate models, we fine tune the parameters to optimize the simulation. A total of 220 simulations have been performed to provide the necessary material for reproducing Cen A's observed properties. A systematic comparison between these models and observations has let us to select four of them that show the best reproduction of Cen A's properties. Parameters of these four models are presented in Table 1. They all have similar angular orbital parameters as well as the initial gas fractions. Most of the variance is in mass ratio, initial scale-lengths, and pericenters, all being major ingredients in merger modelling, which well demonstrate the final result variance from different models. We run simulations using 1.85 million particles per simulation. Additionally, we also run some simulations with 6 and 10 million particles. There was little change with the increased number of particles demonstrating convergence of our simulations. 

\begin{figure*}
\includegraphics[scale=0.6]{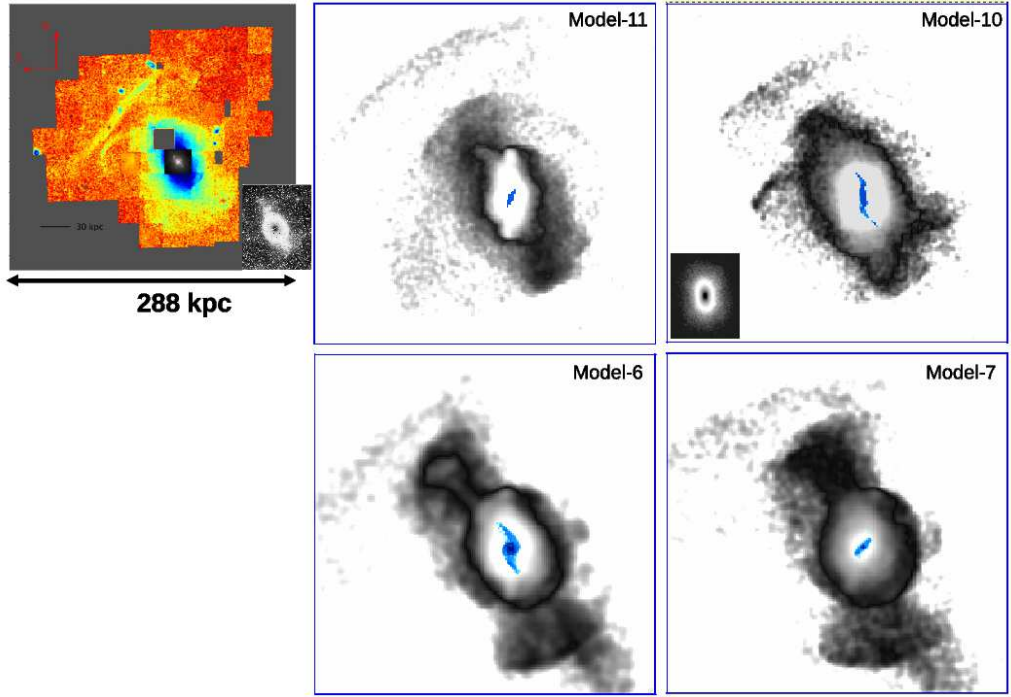}
\caption{Comparison of observed (left) stellar mass distribution with simulations (middle and right panels).
Observational data on the top left panel are from
\citet{Crnojevic2016}. The observed stellar density map has been scaled in physical size consistent with
the simulated Models. The small insert next to the observed map shows the image from D.\ Malin published in \citep{Israel1998}
for comparison with the Model-10 (see the insert in the top-right panel). Four simulation results are shown on the right
panels (see Table 1). For clarity the color in the center of 
each model image has been inverted and overlapped. The final gas morphology has been superimposed with blue color on
each simulation image. The size of each simulated image is 350 by
350 kpc.} 
\label{fig:image} 
\end{figure*}


In what follows, we compare several features from our simulations to the observed properties of Cen A to assess the robustness of our methods and results.

\section{THE RESULTS}

\subsection{Galaxy morphology and faint features formed during merger}

Fig.\ref{fig:image} shows the stellar mass distribution resulting from the four models and compares them to the observations made by \citet{Crnojevic2016} and image from D.\ Malin published in \citet{Israel1998}.  All models reproduce the main features observed in the Cen A imagery. 
In the deepest observations from \citet{Crnojevic2016}, the halo of Cen A shows
an elongated shape along the major axis.
At both ends of the major axis, there are several over-densities and clumps, in particular on the bottom side of the galaxy. This behavior is well reproduced by our simulations and identified to be main residuals of the merger event. 
Interestingly, our simulations are able to reproduce one of the most stunning features in Cen A's halo, i.e., the long stream to the north-east of the galaxy, as a result of the central major merger. The stream was first discovered in the ground-based imaging presented by \citet{Crnojevic2016} and further followed-up with HST \citep{Crnojevic2019}; it is 60 kpc long and located at a projected distance of 80~kpc from Cen A, and it features a clear remnant (dubbed Dw3) which hosts a candidate nuclear star cluster (Seth et al., in prep.). The stream could have originated from the tidal influence of Cen A on a $M_V \sim -15$ dwarf galaxy, leading to the observed S-shape of the stripped remnant's outer isophotes \citep{Crnojevic2016}. However, the shape of Dw3's tidal tails, and in particular their straightness, is somewhat unusual: one possible interpretation is that this is the result of a recent accretion event, which is observed close to apocenter and with a large velocity component in the direction of the plane of the stream (S. Pearson, private communication).
It is intriguing that this feature can be naturally obtained from our simulations: this could possibly imply that Dw3 has a tidal dwarf origin instead \citep{Fouquet2012,Hammer2013,Yang2014,Ploeckinger2015,Baumschlager2019,Kroupa2012,Metz2007}. This scenario can be tested by obtaining kinematics of the remnant to assess the presence/absence of dark matter. We also highlight that one model (Model-10) reproduces as well a second, curved stream that can be seen on the left side of the images in the top-left \citep{Crnojevic2016} and top-right panels of Fig.\ref{fig:image}.

Observations have revealed a warped HI gas disk \citep{Schiminovich1994,
Israel1998,vanGorkom1990}, which is roughly perpendicular to the major axis of the
stellar component. In Fig.\ref{fig:image} the gas component is the blue component overlapping the center of the galaxy. Three models show such an edge-on gas disk component, and two of them, Model-11 and Model-7, have gas disk position angles roughly consistent with observations. We notice however, that the gas disk component presents large angular changes with time, which means that capturing the proper position angle is a difficult and lengthy task. To get a better match, 
more models are needed for fine tuning. 

Perhaps the major weakness of our modeling is the overall size of Cen A, which is systematically larger in simulations than the observed one. 
For example, in Model-10 the distance from the galaxy center to the faint features found in the bottom of the image  is about 22\% larger than that observed, while Model-7 leads to even more overestimated size (60\% larger). Moreover, for Model-10 and Model-7, the projected distance from the galaxy center to the straight
stream is 70 and 100\% larger than that observed, respectively. Thus our modeling is well indicative
of the structures formed during the major merger that has occurred in Cen A, but does not
represent an accurate reproduction of this galaxy, contrary to the one obtained in \citet{Hammer2018} for M31. We notice that to get such an accurate reproduction of M31 required to perform close to a thousand different models, most of them being dedicated for fine tuning. To recover the projected distance between the straight stream and the galaxy center different projected angles need to be considered. Alternatively, the mismatch could be due to the low baryon-to-dark matter ratio adopted here,
since it possibly leads to a merger that is too energetic and expels too large quantities of material in the galaxy outskirts. Among several fine tuning parameters that can be experimented with, one may consider changing the initial scales of the progenitors, as well as the merger pericenter.

\begin{figure*}
\includegraphics[scale=0.6]{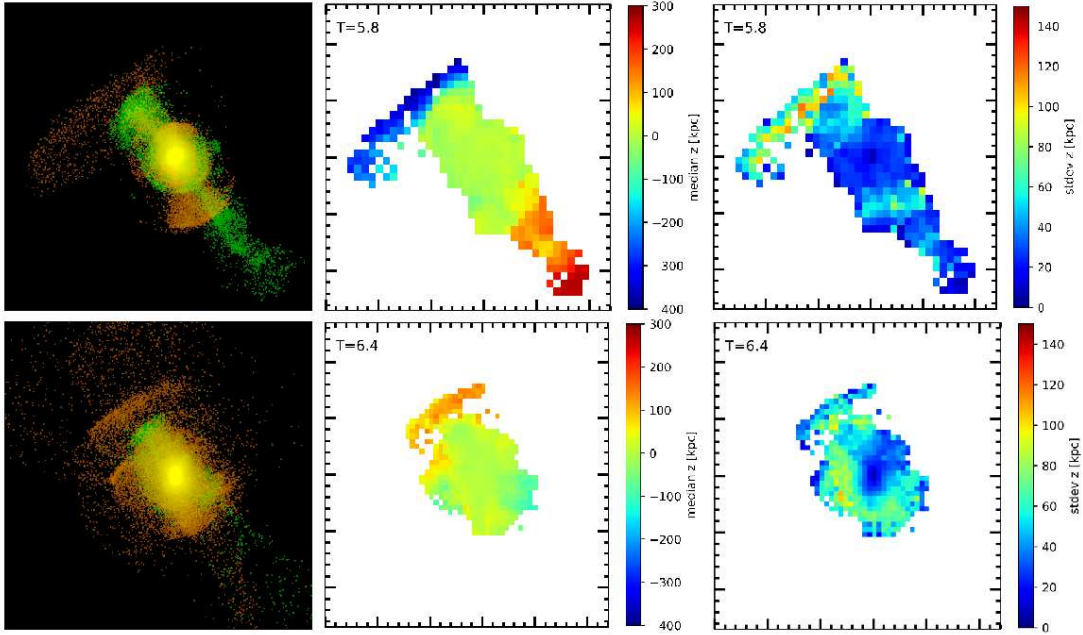}
\caption{The left panels show the particle distribution with red and green
color indicating each of the two progenitors, for Model-6 (top) and Model-10 (bottom), respectively. 
The middle panels indicate the stellar particles distance with respect to the center of galaxy. The 
negative distance values are behind the galaxy and positive values in
front of the galaxy. The right panels show the standard deviation of the distance distribution. Size of each panel is 540 by 540 kpc.} 
\label{fig:Zdis}
\end{figure*}

The left panels of Fig. \ref{fig:Zdis} show how the stellar particles from 
the two progenitors are distributed. While both progenitors contribute to the central regions, there
are clear differences in their outskirt distributions. For example, the straight stream
is always coming from a single progenitor as well as the shell-like structures 
on both top and bottom sides of the galaxy (see red particles). The other progenitor (see green particles) may also contribute to the elongated distribution of the residuals along the major axis, though with large differences from one model to another.

The distance distribution of residuals in the galaxy outskirts is 
shown in the middle panels of Fig. \ref{fig:Zdis}. For Model-6 the straight stream is found 300 kpc behind Cen A. On the contrary in Model-10 it is 100 kpc in front of Cen A.
As shown by \citet[see their table 5]{Crnojevic2019}, the distance estimates along this stream
are not well constrained by observations, with values ranging from 150 kpc in front of Cen A, to 750 kpc further away. Uncertainties from observations add another reason why fine tuning of the model could not be accomplished, for the moment.


In the right panels of Fig. \ref{fig:Zdis}, the color coding indicates the  standard deviation of the distances. In
most cases, the standard deviation is around 60 kpc in the straight stream such as in
Model-10, while it may reach 110 kpc in Model-6. The observed difference in
distance at different positions along the straight stream is quite large reaching up to
900 kpc in \citet{Crnojevic2019} (between Dw3-WFC3 and
Dw3S-ACS). This is significantly larger difference than the distance deviation in the models, and
may indicate that the straight stream is made of different components and/or has a different origin, as discussed above. 

\begin{figure*}
\includegraphics[scale=0.53]{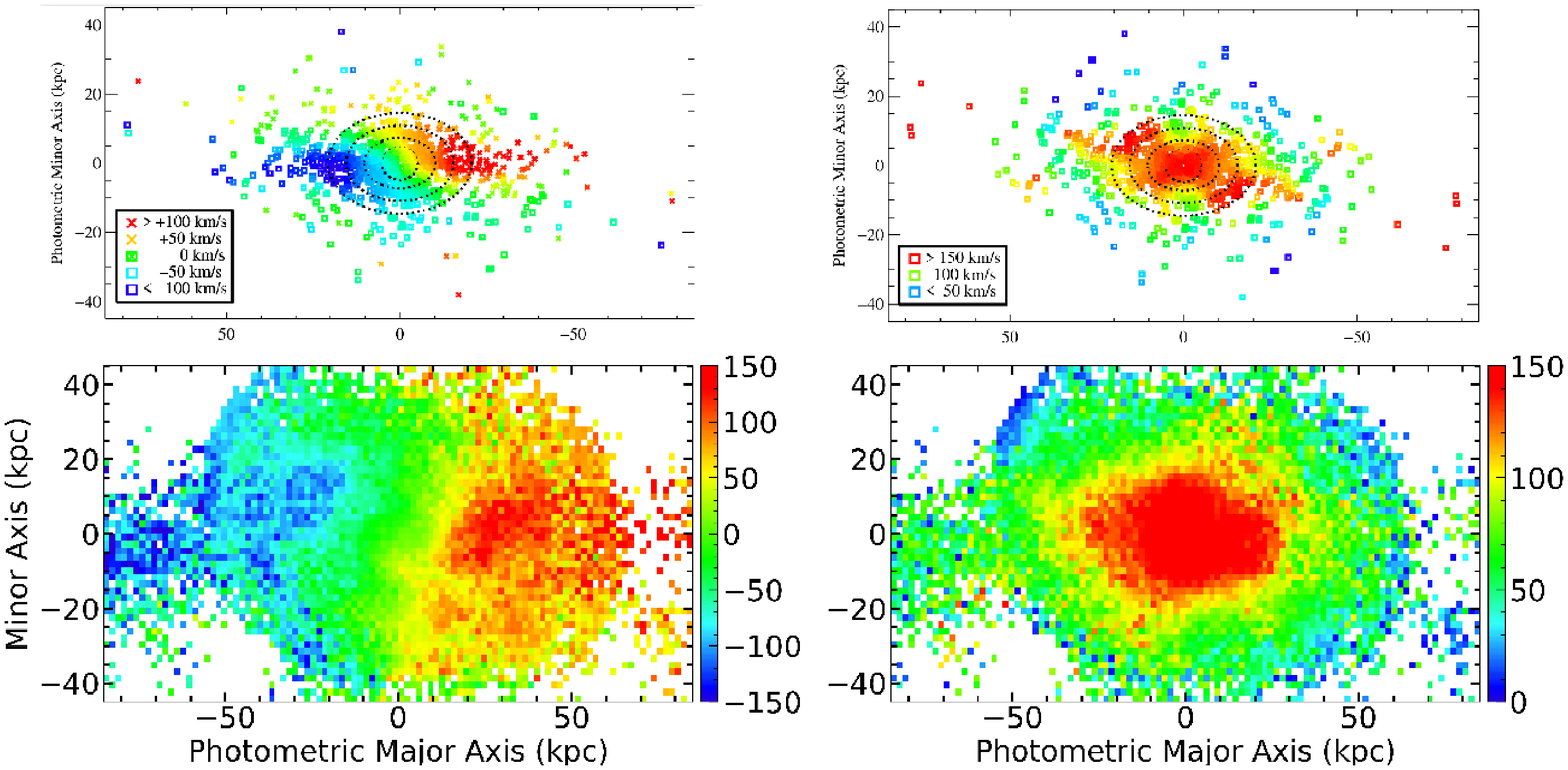}
\caption{Comparison of the observed velocity field and velocity dispersion (top panels from \citealt{Peng2004PN} measurements of PN) to that from Model-6 (bottom panels). Line of sight velocities are shown on the left, while the right column shows the associated velocity dispersion. The same color map is used for observations and simulations.
}
\label{fig:VF}
\end{figure*}

\subsection{Velocity Field}

Over the last a few decades, the kinematic properties of Cen A have been
studied using different tracers, including spectroscopy of integrated stellar light
\citep{Wilkinson1986}, of PN \citep{Hui1995, Peng2004PN}, and of GCs
\citep{Peng2004GCS, Beasley2008, Woodley2007, Woodley2010kinematic}. These studies provide valuable
constraints on the kinematic properties of Cen A and total mass.

Bottom panels of Fig.\ref{fig:VF} show the velocity field (left) and dispersion map (right) for Model-6. In
this figure, the galaxy has been rotated to have the major axis 
along to the x-axis to ease the comparison with the observations by
\citet{Peng2004PN} shown in the top two panels.
\citet{Peng2004PN} used PN to study the velocity field. The final Model-6 remnant shows a rotation along the major axis
with an amplitude of about 150 km/s, which is fully consistent with the observations
\citep{Peng2004PN}. The central peak in the velocity dispersion map
reaches about 150 km/s, which is also consistent with observations
\citep{Peng2004PN}. Given the discrete tracer (PN) the observations show a sparser sampling than that of the models.

The line of zero velocity traced by \citet{Peng2004PN} is both misaligned
and twisted with respect to the photometric axes \citep{Peng2004PN}. The
simulated line of zero velocity shows similar features but with some geometrical differences.

\subsection{Total mass distribution}

\begin{figure*}
\center
\includegraphics[scale=0.9]{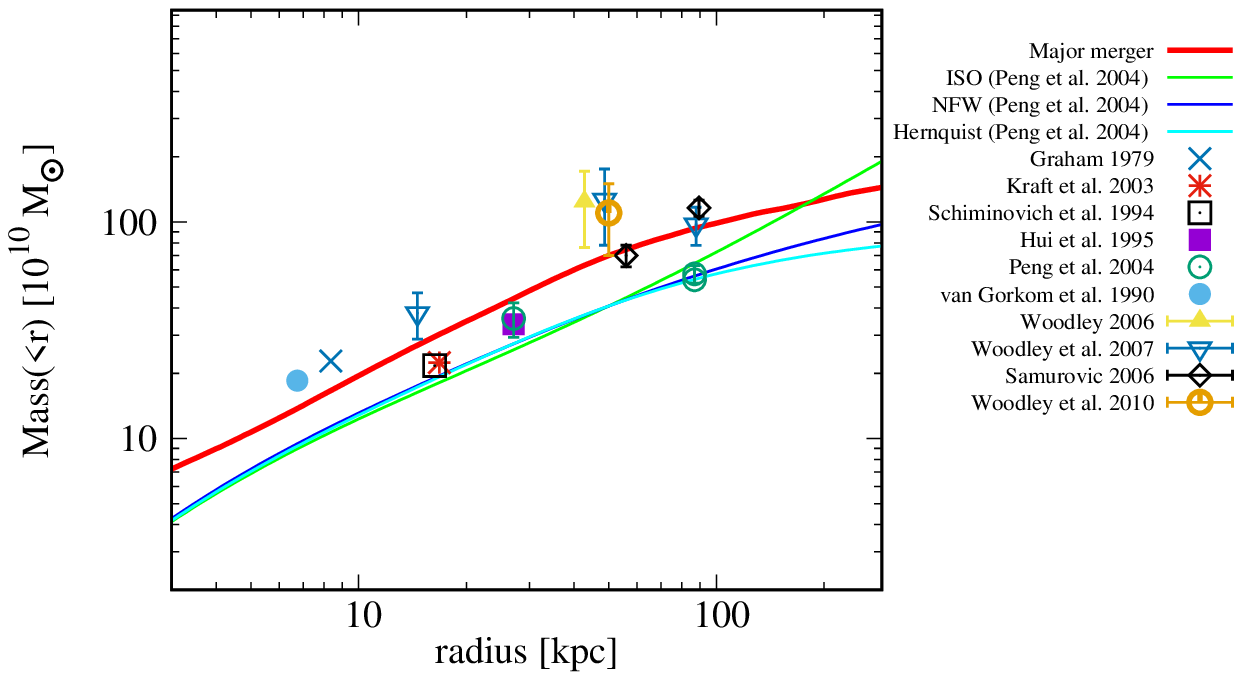}
\caption{A comparison of total mass distribution of Model-6 (red solid line) to values from literature as listed on right side of the Figure.} 
\label{fig:Mass}
\end{figure*}

Measurements of total mass of Cen A are important because they reveal the 
amount of dark matter.  Several different methods and tracers have been used to estimate the total mass profile for Cen A. Their values, compiled from the literature and scaled to the galaxy distance of 3.8~Mpc are plotted with different symbols and lines as shown on the right side of Fig. \ref{fig:Mass}.  The total mass tracers include GCs \citep{Woodley2006, Woodley2007, Woodley2010kinematic}, PN 
\citep{Peng2004PN, Hui1995, Samurovic2006}, HI gas \citep{Graham1979, vanGorkom1990, Schiminovich1994}, and X-ray emission \citep{Kraft2003}. There is 
a significant scatter between
the total mass measurements at comparable radii with a clear outward increase indicating that the galaxy is not dominated by dark matter in the inner $\sim 5$~R$_\mathrm{eff}$ \citep[see also][for further discussion]{Peng2004PN}. The red line (Major merger) Model-6 simulation follows the outward increase in the total mass and lies in between observed data points.  

\subsection{Stellar mass profile}

\begin{figure} 
\includegraphics[scale=1.0]{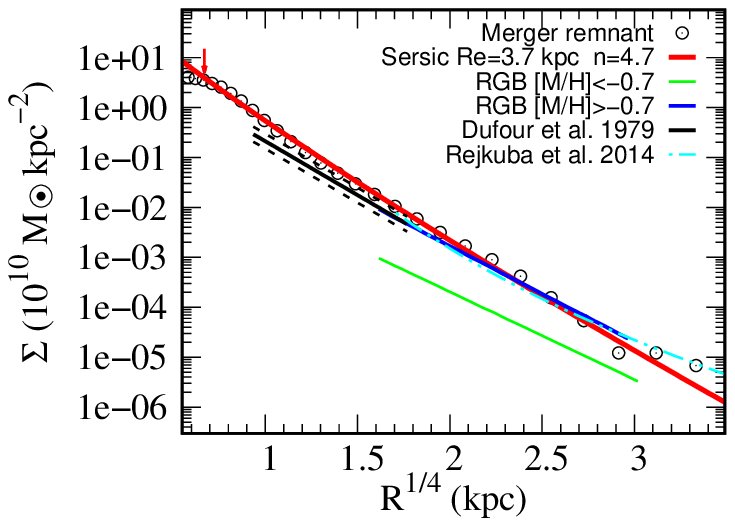} 
\caption{A comparison of the stellar mass distribution derived from observations with that of Model-6, which is fitted with a S{\'e}rsic profile (red line, see parameters indicated on the top-right).
The red arrow indicates the two times softening length within which model predictions are limited by resolution. Fitting parameters 
for all of the four models are shown in Table 1. The
observed surface brightness profile from \citet{Dufour1979} converted to
stellar mass surface density (see the text) is shown with black solid line  and black dashed-lines
indicate the effect of a 0.15 dex systematic offset due to IMF variations. The observed surface 
brightness of RGB stars from \citet{Bird2015} are scaled to be consistent with that of \citet{Dufour1979}  and shown with blue and green lines for metal-rich and metal-poor population, respectively. The cyan line shows the power law distribution of light measured from RGB star counts from \citet{Rejkuba2014}, that was motivated by excess of light along the major axis for fields beyond $\sim 50$~kpc.} 
\label{fig:SB} 
\end{figure}

\citet{Dufour1979} have measured the surface brightness profile in the
central region in $B$ band and obtained a very good fit with a de Vaucouleur's law. The extended surface brightness of Cen A to its outskirts has been measured using observations of individual RGB stars in the halo \citep{Crnojevic2013, Rejkuba2014,Bird2015}. \citet{Crnojevic2013}  and \citet{Bird2015} resolved stellar halo of Cen A using VIMOS@VLT. \citet{Crnojevic2013} observed two fields along the north-eastern major axis and another two fields along south-east minor axis, spanning a range of distances between $\sim 40-80$~kpc. \citet{Bird2015} had a single VIMOS pointing at $\sim 65$~kpc, but they combined their results with the RGB star counts and MDF measurements from previous Hubble Space Telescope $(HST)$ data \citep{Harris1999, Harris2000, Harris2002, Rejkuba2005} to derive the density falloff independently for metal-rich
and metal-poor stars between $\sim 8-65$ kpc. They found that both the
metal-rich and metal-poor populations can be well fitted with a de Vaucouleur's law profile. \citet{Crnojevic2013} noted a higher RGB number density along the major axis that deviated significantly from a de Vaucouleur's profile beyond $R \geq 75$~kpc. Their data had insufficient coverage to distinguish between the genuine flattening of the radial profile and a presence of small scale substructures in the halo. \citet{Rejkuba2014} used ACS and WFC3 cameras on board the HST to extend resolved stellar halo studies out to a projected distance of 140 kpc along the galaxy major axis and 90 kpc along the minor axis. They found that in the outer halo, beyond $\sim$10 R$_{\rm eff}$, there is a systematic increase in number counts along the major axis, even after accounting for field-to-field variations due to small scale substructures. A de Vaucouleur's profile was shown to provide a good fit in the inner part of the galaxy as well as along the minor axis, while a power law provided a better fit along outer halo major axis. 

Fig.\ref{fig:SB} shows the surface mass density distribution of the Model-6 remnant, including its fit with a S{\'e}rsic profile. The S{\'e}rsic
index is 4.7, which is consistent with the observed de Vaucouleur's profile
($n=4$). For comparison we plot in Fig.\ref{fig:SB} the  \citet{Dufour1979} $B$ band profile converted to stellar mass
surface density. To convert the $B$ surface brightness to a stellar mass surface density, we have used the $B-V$ color vs.\ stellar mass to light ratio relation \citep{Bell2003}. A $B-V$ color of 0.84 is adopted from \citet{Dufour1979}, and $V_{\sun} = 4.82$ from  \citet{Bell2001}. 

The surface brightness distributions
of RGB stars have been re-scaled to that of \citet{Dufour1979}. In this figure we also compare the simulation with
the surface brightness profile of RGB stars from \citet{Bird2015}. In cyan we show the power law fit to RGB stars surface distribution from \citet{Rejkuba2014} rescaled to surface mass density as above. The excess of light in the outermost bins of the simulated mass surface density is well matched with this power law.
The slope of the surface brightness and the stellar surface number counts can be robustly used for comparison to the model, because we are confident
that the total stellar mass used in the models is well consistent with
observations (see section 4.3).
The  observed surface brightness profiles from integrated light and RGB star counts are consistent with profiles of the four merger remnants examined here, for which the S{\'e}rsic index range from 4.4 to 6.2, i.e.\ slightly larger than observations. 
Indeed the profile from \citet{Dufour1979} is slightly shallower (Fig.\ref{fig:SB}). This could be either due to IMF variations (see black dashed lines showing a 0.15 dex systematic offset) or due to an increasing extinction in the central region. 

The half light radius of Cen A is changing with wavelength. 
\citet{Dufour1979} estimated it in $B$ band to be
305${\arcsec}$ (5.6 kpc) with a de  Vaucouleurs profile. The half mass
radius for the four models are ranging from 3.7 to 6.3 kpc as shown in Table 1. 

\subsection{Star formation during the major merger}

\begin{figure}
\includegraphics[scale=0.50]{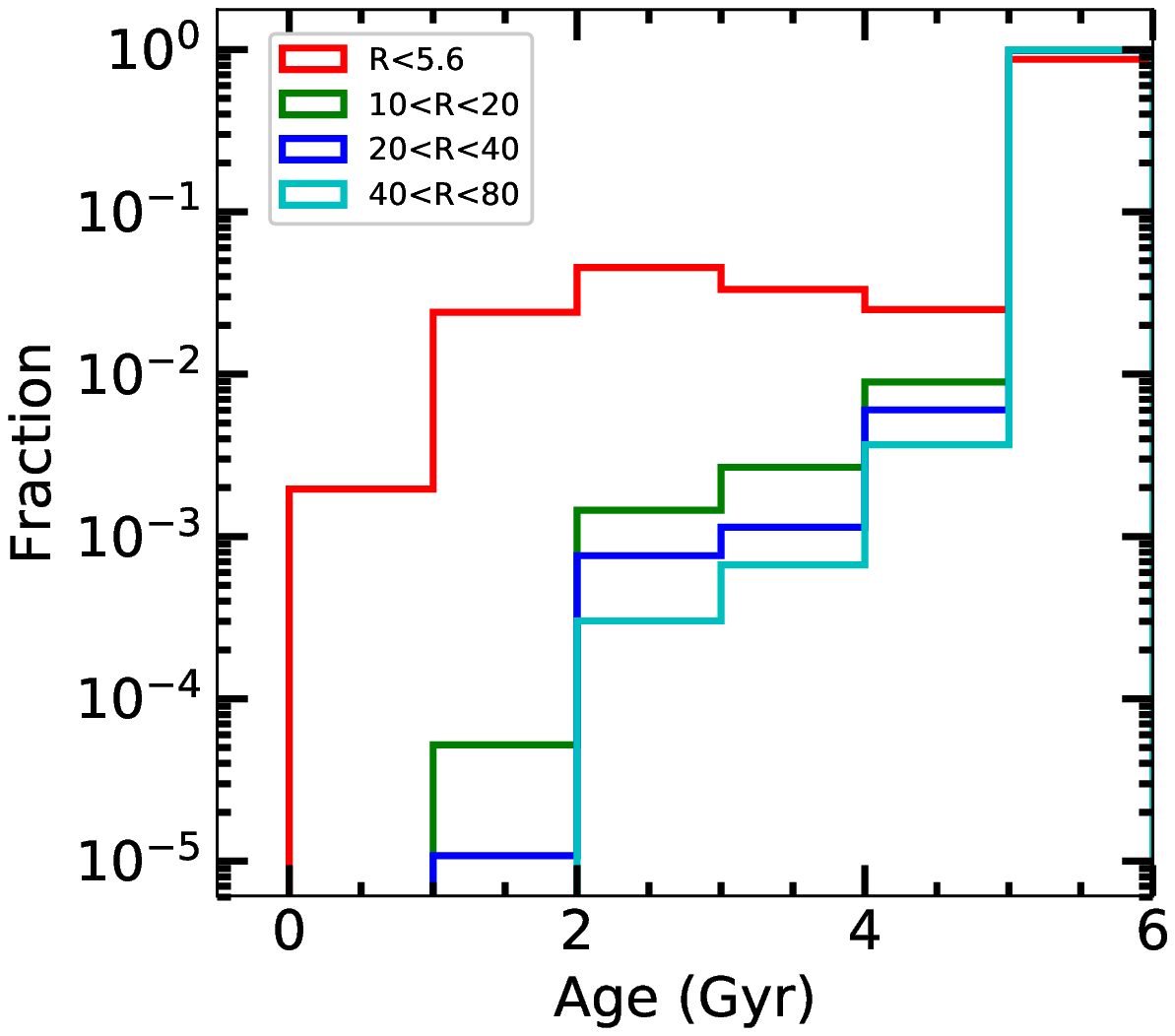}
\caption{The stellar age distribution for stars at different radii based on Model-6. As the simulation started 6 Gyr ago, the peak at 6 Gyr is artificial and represents stars with a minimum age of 6 Gyr.  Within 5.6 kpc there is a peak around 2 Gyr, which likely indicates the fusion epoch during which star formation occurred in a central starburst (see, e.g., \citealt{Hammer2018}).  }
\label{fig:Age}
\end{figure}

In Fig.\ref{fig:Age} the age distribution of stars at different radii is
shown. However, since the simulation is started 6 Gyr ago, the peak at 6 Gyr is artificial and indeed
represents stars with ages larger or equal to 6 Gyr.
This could have been corrected by re-sampling the age of
these stars  according to the observational constraints, though such considerations are out of the scope of this paper.

\citet{Rejkuba2011} found that the halo contains
$20-30\%$ of stars that have been formed $2-4$ Gyr ago in a field located $\sim$~40 kpc south of the galaxy center, which is also consistent with the
age distribution of global clusters \citep{Woodley2007, Woodley2010}.  The latter
studies found many young GCs in the halo, with youngest ages
around 2 Gyr, which confirms our choice of a recent fusion epoch. The above results are consistently reproduced by our modeling (see Fig.\ref{fig:Age}) after considering radii ranging from 10 to 80 kpc. 
A larger fraction of recent star formation is expected in the central regions (within the half mass radius, R$_{\rm eff}=5.6$ kpc) after a major merger, during which interaction between progenitors and then fusion have enhanced gas compression and favoured its transformation into stars.



\subsection{Metallicity distribution and radial metallicity gradient}

\begin{figure}
\includegraphics[scale=0.65]{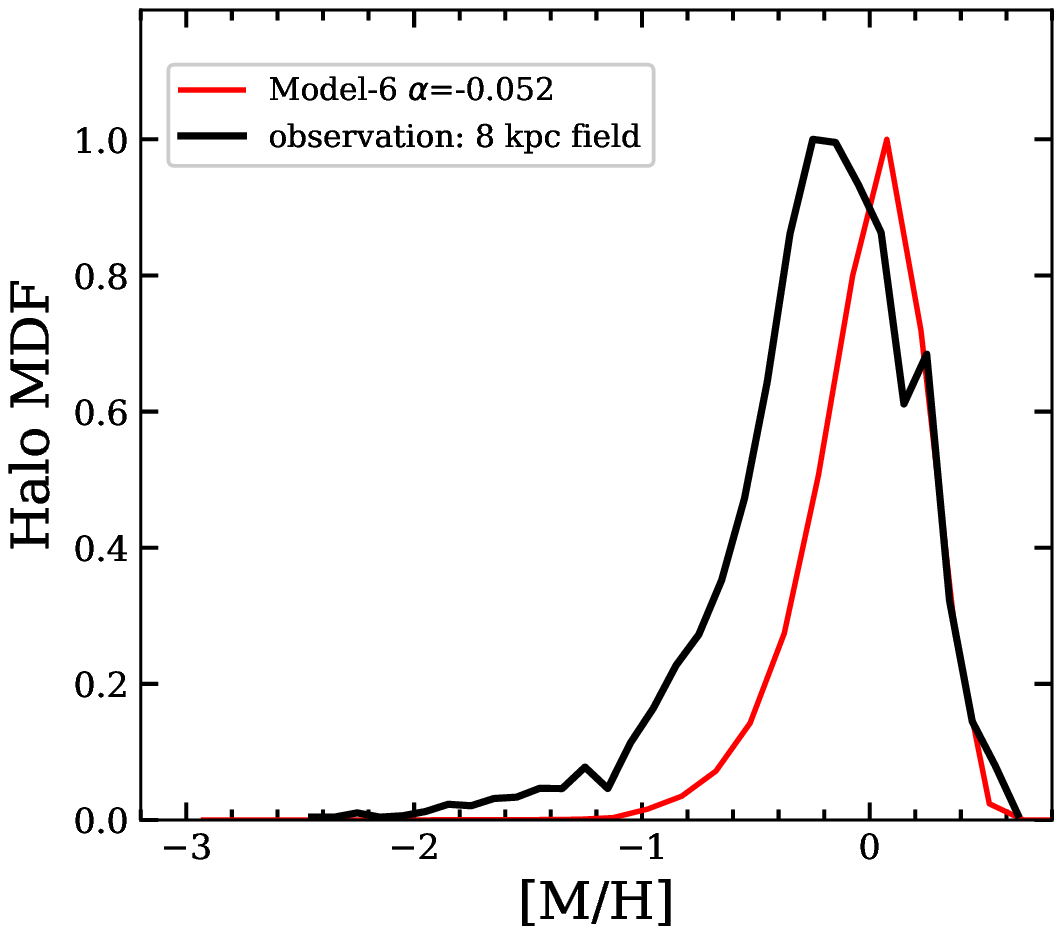}
\includegraphics[scale=0.65]{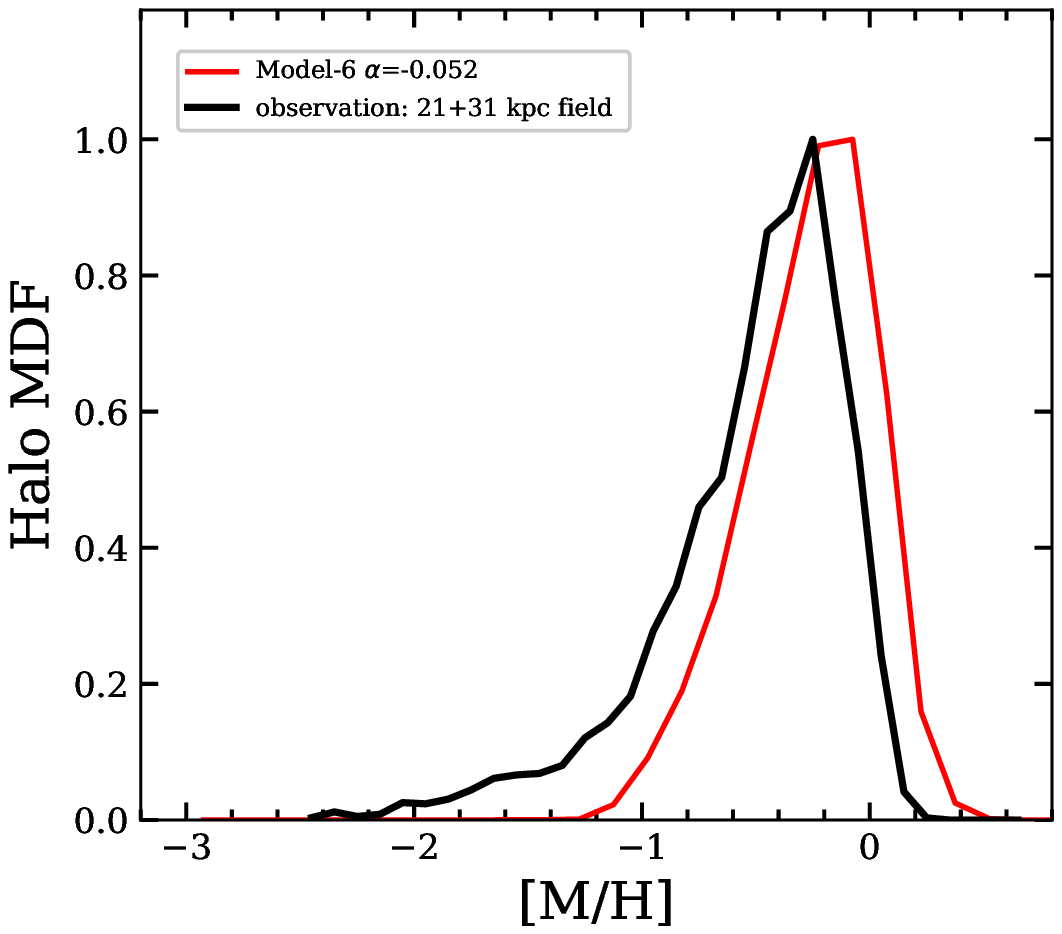}
\caption{Metallicity distribution function compared with observations 
\citep{Harris1999,Harris2000,Harris2002}. {\it Top panel:} field with projected
distance of 8 kpc to center of Cen A. {\it Bottom panel:} field with projected
distance of 20-30 kpc to the center of Cen A. To extract stellar metallicity from the models, we have used initial metallicity gradients for the
progenitors, one with ${\alpha}=-0.052$ (see red lines).}
\label{fig:MDF}
\end{figure}

\begin{figure}
\includegraphics[scale=0.55]{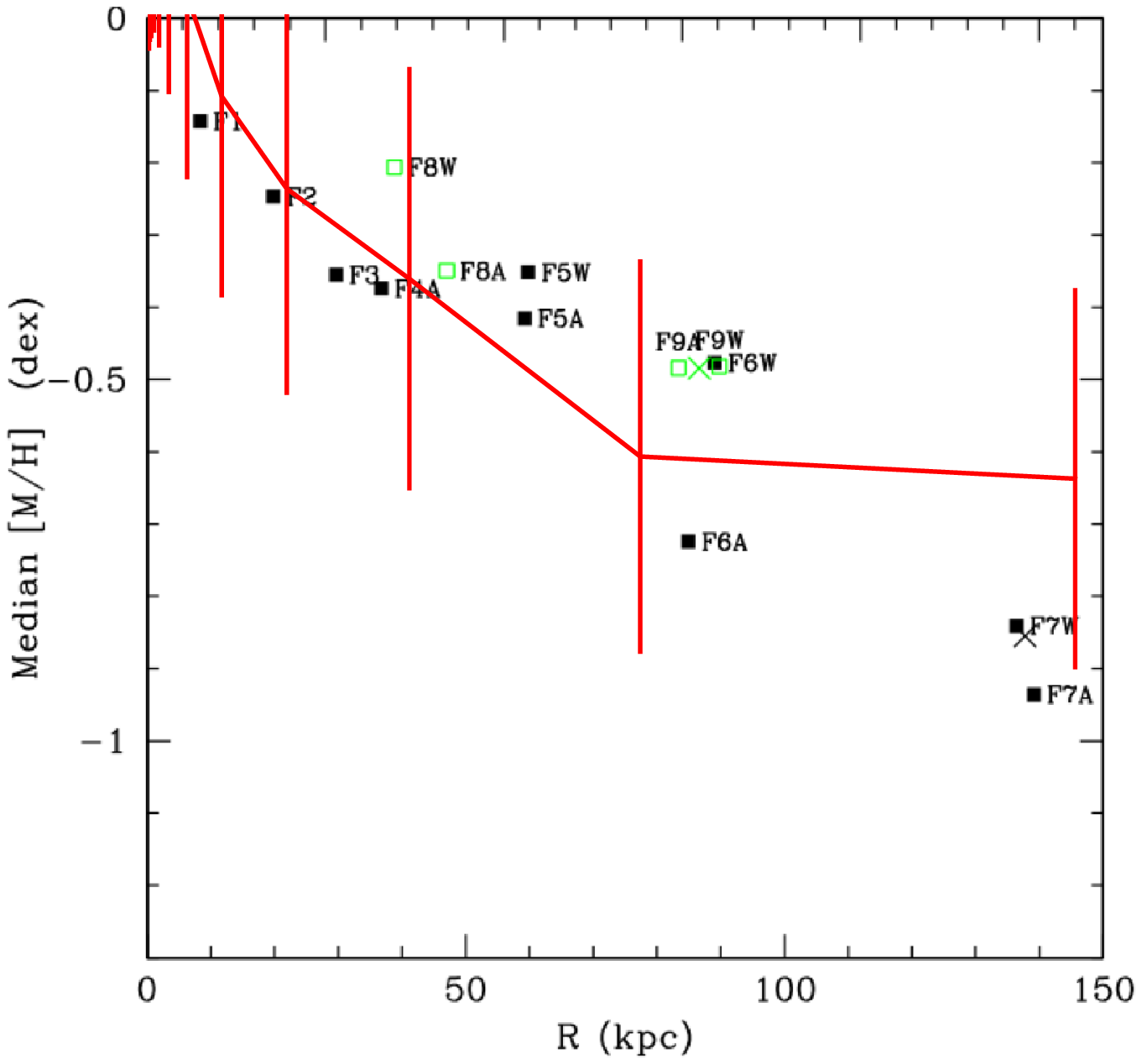}
\caption{Comparing the mean metallicity gradient between Model-6 (red line) and observation data from \citet{Rejkuba2014} (points).  The red-line is the mean value for each bin, and the error bars show the scatter of metallicity in each bin in the model.}
\label{fig:RMH}
\end{figure}

Thanks to the proximity of Cen A, its metallicity content can be measured from
resolved stars. Metallicity distribution functions (MDFs) of halo stars in Cen A have been determined from the HST photometry 
 \citep{Harris1999,Harris2000,Harris2002} in three fields at projected
distances of 8, 21, and 31 kpc from the galaxy center. They can be used to constrain the merger history of Cen A. Following \citet{Bekki2003}
we have compared the MDF of Model-6 with observed MDFs. For the inner halo
region field (8 kpc), we have selected stars with projected distance to the center
between 7 and 9 kpc.  For comparison to the two outer halo fields (21 and 31 kpc), we have selected stars in the models with projected
distance between 20 to 30 kpc. For the initial metallicity distribution in 
the progenitors we have assumed:

$\rm [M/H] = [M/H]_{R=0} + {\alpha}{\times} R$


The initial metallicity gradient ${\alpha}$ for the progenitors is  
-0.052 dex kpc$^{-1}$, which is from the results of MW open cluster from APOGEE 
and GALAH and consistent with the literature value ranged from -0.035 
to -0.1 dex kpc$^{-1}$ \citep{Carrera2019, Friel1995} 
The central metallicity value $\rm [M/H]_{R=0}$ is set to +0.5, which is very close to the maximum metallicity of MW bulge
\citep{Sarajedini2005, Zoccali2017}.  This value is slightly lower than the maximum
metallicity of M31 bulge, which is around +0.9 (see Fig. 6 of
\citealt{Sarajedini2005}).

Fig.\ref{fig:MDF} compares the MDF for the inner halo field (8 kpc), and for the outer halo (21 and 31 fields).  As pointed out by \citet{Rejkuba2014} the metallicity calibration in \citet{Harris1999,Harris2000,Harris2002} has an offset of 0.2 dex with respect to that in \citet{Rejkuba2014} due to use of different stellar evolutionary models. Also there is a relatively large incompleteness correction at the high metallicity end. Therefore, we have artificially shifted the black line (obeserved MDFs) in Fig.\ref{fig:MDF} by +0.2 dex. 
The median of observed MDFs are still lower than the model for both inner field and out field. These may reflect the incompleteness in photometry for the most metal-rich, reddest and faintest stars.

In Fig.\ref{fig:RMH} we compare the observed metallicity distribution along the radial direction with that from our model. Within 100 kpc, the model will fit the observation data.  At 150 kpc the metallicity is about 0.2 dex higher 
than that from observation, which may need more observation at that scale to 
overcome the variance from field to field.


\subsection{Ongoing star formation in the disk}

\begin{figure}
\includegraphics[scale=0.4]{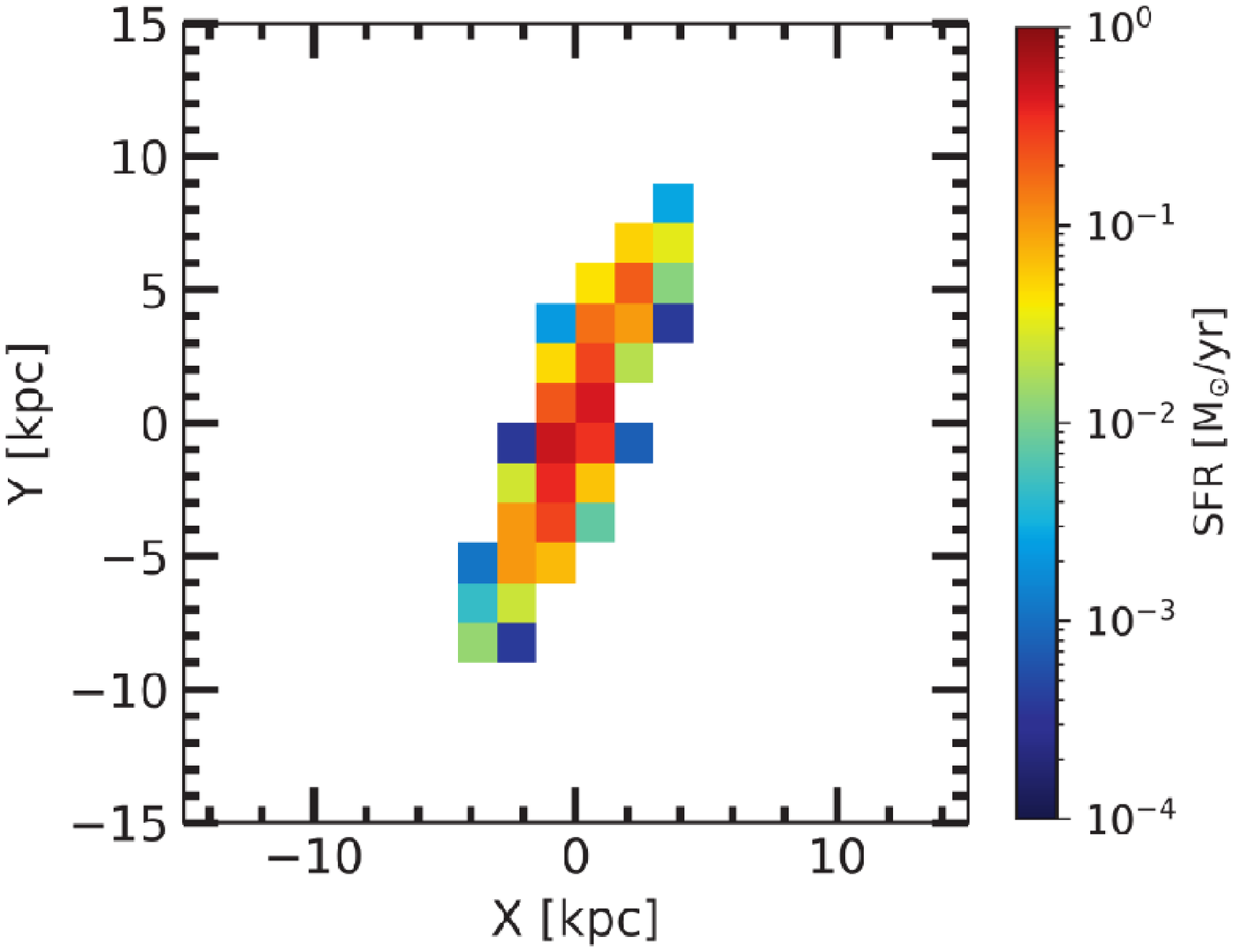}
\includegraphics[scale=0.4]{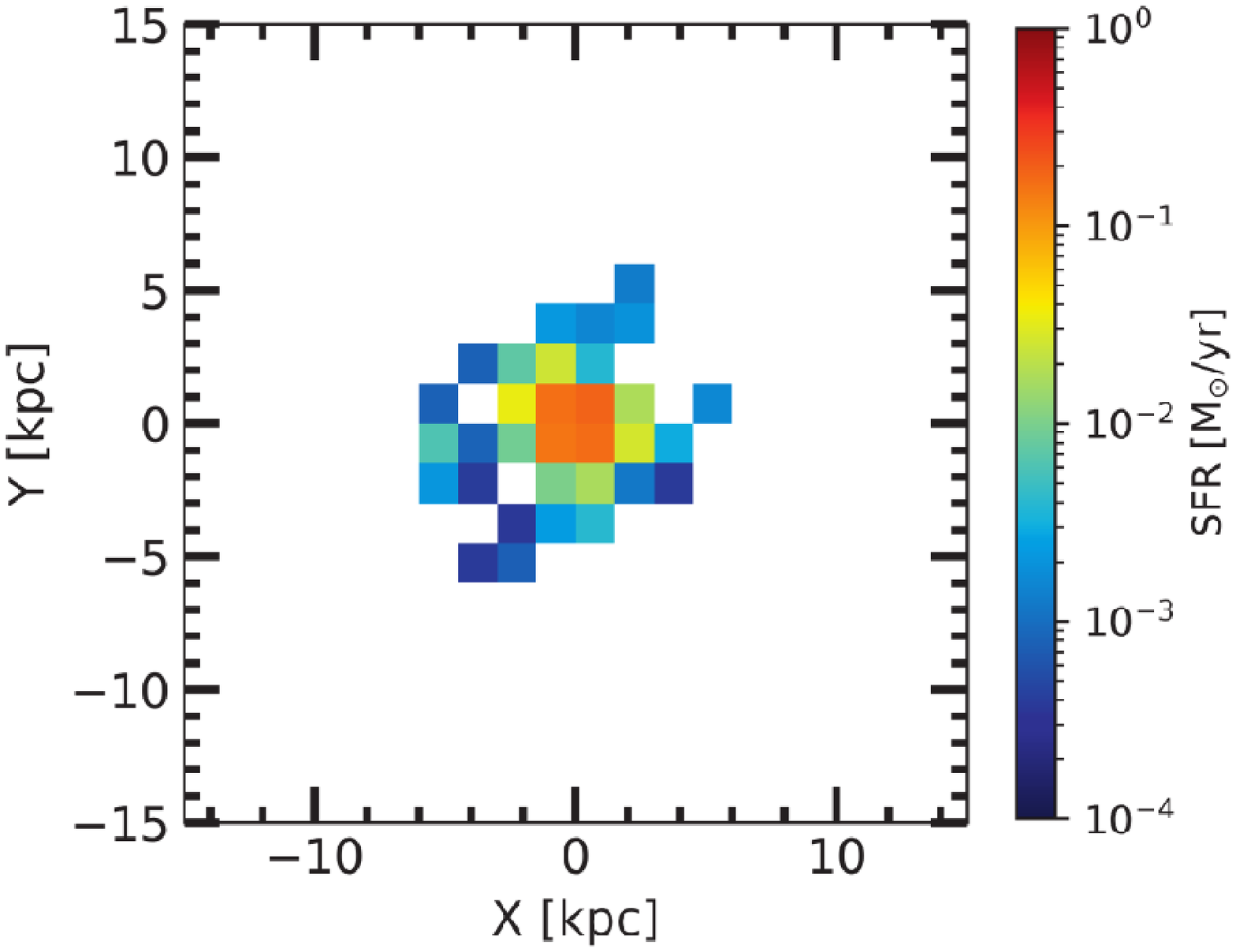}
\caption{The star formation map of Model-11 is shown in the top panel and Model-6 map is in the bottom panel.}
\label{fig:SFR}
\end{figure}

There is still residual gas and star formation in the very centre of Cen A as visible in Fig.\ref{fig:image}
both from observations (gas fraction of $\sim 1$\%) and modeling (3.2 to 8.8\% from Table 1). This is linked
to the central warped HI disk. Both young stars, Wolf-Rayet stars and red supergiants, as well as HII regions have been detected around the dusty disk \citep{Graham1979, Moellenhoff1979, Minniti2004, Kainulainen2009}. The star forming warped disk is also well traced by the $Spitzer$ IRAC map at $8\mu$m \citep{Quillen2006,Quillen2008}, which is well known to trace the star formation \citep{Wu2005}. Furthermore, there is a reservoir of molecular gas fueling low level star formation \citep{Eckart1990, Espada2019}. Possible discrepancy in gas fraction between observations and simulations could be solved by implementing a less efficient feedback in the
modeling to better exhaust the HI gas.

Figure~\ref{fig:SFR} shows the map of star formation rate (SFR) for Model-11 (top) and Model-6 (bottom) simulations. The ongoing star formation confined to the disk is fueled by the remaining gas that has sunk to the center after the merger ended $\sim 2$~Gyr ago. The total SFR estimated for Model-6, -11, -7, and Model-10 are 0.85, 3.5, 1.3, and 0.6 M$_\odot$~yr$^{-1}$, respectively. A recent study by \citet{Espada2019} using ALMA CO (1--0) observations produced high angular resolution ($1\arcsec$) maps towards the dust lane of Cen A. They found a total molecular gas mass of $1.6 \times 10^9$~M$_\odot$, which is higher than previously measured, and derived a star formation rate of $\sim 1$~M$_\odot$~yr$^{-1}$. This is remarkably similar to the values we estimate from our simulations. The ALMA map has a much higher resolution and shows a more structured distribution of SFR surface density than can be inferred from the present simulations. It presents a peak in the central circumnuclear disk, which is reminiscent of what we see in Model-6, albeit on a much smaller spatial scale.

\section{DISCUSSION AND CONCLUSION}

We presented the first hydrodynamical simulations tailored to model the nearest gE Cen A, assuming
a major merger with a mass ratio up to 1.5. From our modeling, the merger event happened 
$\sim 6$ Gyrs ago with the first passage at $\sim 5$~Gyr ago and the fusion of the two progenitors completing 2 Gyr ago. The age distribution of the stars brought in by the progenitors and formed during the merger event is consistent with the stellar age distribution in the halo \citep{Rejkuba2011}.  
With low gas fractions (20\%-40\%) in the progenitors and a small mass
ratio (${\leqslant}$1.5), the merger remnant stellar mass distribution follows a de Vaucouleurs profile, consistent with observations \citep{Dufour1979}.

In the current model the halo region is dominated by stars coming from the
progenitors of the major merger.  The two progenitors have been assumed to be massive spirals, so
it is expected that relatively metal-rich stars spread throughout the halo region.
This may explain the observations showing the relatively high average metallicity of halo stars ${\rm [M/H]} >-1$ \citep{Harris2000, Harris2002, Rejkuba2005} and the shallow metallicity gradient \citep{Rejkuba2014}.

\citet{Malin1978} and \citet{Malin1983} reported the discovery of many faint narrow stellar shells surrounding Cen\,A. Additional shells in the inner parts of the galaxy were uncovered by \citet{Peng2002}, who applied the unsharp masking technique to CCD images of Cen\,A. Such faint shells and filaments were found to be quite common features of many nearby early-type galaxies \citep{Malin1980}. There are two principal scenarios in the literature explaining the shell formation. One assumes accretion of a smaller (spiral) galaxy onto an elliptical. \citet{Quinn1984} showed that shells can be formed by the "phase-wrapping" of dynamically cold material with accreted companion on radial
orbit. The shells can also be created by "spatial wrapping" of debris from thin
disk \citep{Hernquist1988,Dupraz1986}.  The second scenario proposed by
\citet{Hernquist1992} assumes a major merger of two equal mass spirals that results in an elliptical with 
shells. This scenario solves many difficulties with the shell formation in the minor merger
model. In a recent study of the incidence and formation processes of shell galaxies based on the Illustris hydrodynamic cosmological simulation \citet{Pop2018} reported that shell galaxies observed at z=0 preferentially formed through mergers with relatively major merger ($\ga 1:10$ in stellar mass ratio). Our current work where shells are result of a major merger is consistent with the latter scenario, 
and also in agreement with recent simulations by \citet{Bekki2003} and
\citet{Bekki2006}.

Our modeling has some limitations and does not provide a detailed description of all the observed Cen A properties. For example, the simulated residuals that mostly lie along the main galaxy axis occupy a larger area than that found in observations, and not all the geometrical angles are reproduced together. Some improvements are expected either from observations, e.g., a more accurate determination of stream distances, or from modeling, e.g., by fine-tuning parameters and also by considering different baryonic fractions. 

However, the success in modeling Cen A's properties indicates that there are still giant elliptical galaxies that are formed through major mergers in the last few Gyr.  The role of major and minor mergers in the mass assembly of massive spiral and elliptical galaxies had been vastly discussed during the last decades. Due to their lower impact and their longer duration \citep{Jiang2008}, minor mergers are considerably less efficient to activate a starburst, and to distort morphologies and kinematics, or they do it in a somewhat sporadic way \citep{Hopkins2008}. On the other hand, the argument can be counter balanced if dwarf galaxies are as numerous as predicted by $\Lambda CDM$ cosmological models. Observations of moderately distant galaxies ($z_{median}$= 0.65), have revealed that 6 Gyr ago, mass ratio $<$ 5 mergers were quite common \citep{Hammer2005,Hopkins2008,Hammer2009} and similar to expectations from $\Lambda CDM$ \citep{Puech2012}. 

The above could make both M31 and Cen A exceptional since they have experienced a much more recent major merger, about 2 Gyr ago. This also implies a quite efficient mass assembly through mergers in the Local Universe, an evidence that could be further supported if M81 is likely to experience soon a merger with M82 \citep{Smercina2019}. 

Perhaps even more intriguingly, 
\citet{Muller2018,Muller2019} showed that dwarf galaxy satellites surrounding Cen A are corotating within a gigantic thin plane that has a measured rms height of 133 kpc and a semi major axis length of 327 kpc. \citet{Woodley2006} compared the kinematics from $\sim 340$ Cen A GCs with over 60 satellite members of Cen A and nearby M83, finding similarities in rotation amplitude, rotation axis, and velocity dispersion between the halo of Cen A and the Centaurus group as a whole. The plane of satellites reported in \citet{Muller2018} is aligned with the galaxy major axis and satellites on the north-east side of Cen A are approaching, while those on the south-west side are receding, indicating a coherent rotation in the same direction as PNe \citep[see Fig.~1 in][]{Muller2018}. M31 shares a similar alignment of the gigantic plane with the Giant Stream \citep{Hammer2013,Hammer2018}, and both planes are aligned with the line of sight. Such gigantic plane structures are not well understood in $\Lambda CDM$ cosmology \citep{Pawlowski2014}, which calls for further observational and simulation studies of Cen A and of its dSph satellites.

\section*{Acknowledgments}

The authors thank the referee for helpful suggestions.
The computing task was carried out on the HPC cluster at China National
Astronomical Data Center (NADC). NADC is a National Science and Technology
Innovation Base hosted at National Astronomical Observatories, Chinese Academy
of Sciences.  This work has been supported by the China-France International
Associated Laboratory 'Origins'. 
Research by DC is supported by NSF grant AST-1814208.

{\it{Data availability: The data underlying this article will be shared on reasonable request to the corresponding author.}}

\label{lastpage}

\bibliographystyle{mn2e}
\bibliography{mn}

\end{document}